# On the Connectivity and Multihop Delay of Ad Hoc Cognitive Radio Networks

Wei Ren,      Qing Zhao*,      Ananthram Swami

## Abstract

We analyze the multihop delay of ad hoc cognitive radio networks, where the transmission delay of each hop consists of the propagation delay and the waiting time for the availability of the communication channel (*i.e.,* the occurrence of a spectrum opportunity at this hop). Using theories and techniques from continuum percolation and ergodicity, we establish the scaling law of the minimum multihop delay with respect to the source-destination distance in cognitive radio networks. When the propagation delay is negligible, we show the starkly different scaling behavior of the minimum multihop delay in *instantaneously connected* networks as compared to networks that are only *intermittently connected* due to scarcity of spectrum opportunities. Specifically, if the network is instantaneously connected, the minimum multihop delay is asymptotically independent of the distance; if the network is only intermittently connected, the minimum multihop delay scales linearly with the distance. When the propagation delay is nonnegligible but small, we show that although the scaling order is always linear, the scaling rate for an instantaneously connected network can be orders of magnitude smaller than the one for an intermittently connected network.

This work was supported in part by the Army Research Laboratory under Grant DAAD19-01-C-0062 and the NS-CTA Grant W911NF-09-2-0053, and by the National Science Foundation under Grants ECS-0622200 and CCF-0830685. Part of this work was submitted to 2010 *ICC*.

W. Ren and Q. Zhao are with the Department of Electrical and Computer Engineering, University of California, Davis, CA 95616. A. Swami is with the Army Research Laboratory, Adelphi, MD 20783.

* Corresponding author. Phone: 1-530-752-7390. Fax: 1-530-752-8428. Email: qzhao@ece.ucdavis.edu

# I. INTRODUCTION

The basic idea of opportunistic spectrum access is to achieve spectrum efficiency and interoperability through a hierarchical access structure with primary and secondary users [1]. Secondary users, equipped with cognitive radios [2] capable of sensing and learning the communication environment, identify and exploit instantaneous and local spectrum opportunities without causing unacceptable interference to primary users [1].

In this paper, we focus on the connectivity and multihop delay of ad hoc cognitive radio networks. Due to the hierarchical structure of the spectrum sharing, these issues are fundamentally different from their counterparts in the conventional homogeneous networks. In particular, even in a static secondary network, the communication links are dynamic due to the spatial and temporal dynamics of the primary traffic. As a consequence, the connectivity of the secondary network depends not only on its own topological structure, but also on the topology, traffic pattern/load, and interference tolerance of the primary network. The multihop delay in the secondary network consists of not only the propagation delay but also the waiting time at each hop for the availability of the communication channel, *i.e.,* the occurrence of a spectrum opportunity offered by the primary network. It is this interaction with the primary network that complicates the analysis of the connectivity and multihop delay of the secondary network.

## A. Main Results

Our technical approach rests on theories of continuum percolation and ergodicity by adopting a two-dimentional Poisson model for both the secondary and the primary networks. A disk model for signal propagation and interference is used as a starting point, which allows us to highlight the fundamental interactions between the primary and the secondary networks without delving into potentially intractable details.

We first analytically characterize the connectivity of the secondary network, where the connectivity is defined by the finiteness of the minimum multihop delay (MMD) between two randomly chosen secondary users. Specifically, the network is disconnected if the MMD between two randomly chosen secondary users is infinite almost surely (a.s.). The network is connected if the MMD between two randomly chosen secondary users is finite with a positive probability.

Under the Poisson model, the key parameter that characterizes the topological structure of the secondary network is the density $\lambda_S$ of the secondary users. For a given transmission power and interference tolerance of the primary network, the key parameter that characterizes the impact of the primary network is the density $\lambda_{PT}$ of the primary transmitters that represents the traffic load of the primary network. The connectivity of the secondary network can thus be characterized by a partition of the $(\lambda_S, \lambda_{PT})$ plane as shown in Fig. 1. Specifically, we show that when the temporal dynamics of the primary traffic is sufficiently rich (for example, independent realizations of active primary transmitters and receivers across slots), whether the secondary network is connected depends solely on its own density $\lambda_S$ and is independent of the density $\lambda_{PT}$ of the primary transmitters. In other words, no matter how heavy the primary traffic is, the secondary network is connected, either instantaneously or intermittently, as long as its density $\lambda_S$ exceeds the critical density $\lambda_c$ of a homogeneous network (*i.e.,* in the absence of the primary network). Note that when $\lambda_S > \lambda_c$, there is a.s. a unique infinite connected component in the secondary network formed by topological links connecting two secondary users within each other's transmission range. We further show that for any two secondary users in this infinite topologically connected component, the MDD is finite a.s.

While the secondary network is connected and the MDD is finite with positive probability whenever there are sufficient topological links (*i.e.,* $\lambda_S > \lambda_c$), there may not be sufficient *communication links* to make the network *instantaneously connected* at any given time. The latter is determined by the traffic load of the primary network. As illustrated in Fig. 1, for any given density $\lambda_S$ of the secondary network with $\lambda_S > \lambda_c$, there exists a maximum density $\lambda_{PT}^*(\lambda_S)$ of the primary transmitters beyond which the secondary network is only *intermittently connected*. When intermittently connected, the secondary network has no infinite connected component formed by communication links at any given time. Messages can only traverse the topological path connecting two secondary users by making stops in between to wait for spectrum opportunities.

It is thus natural to expect that the MDD will behave differently in an instantaneously connected secondary network as compared to an intermittently connected secondary network. Indeed, we show in this paper that the scaling behavior of the MDD with respect to the source-destination distance is starkly different depending on the type of the connectivity. To highlight the impact of the waiting time for spectrum opportunities on the MMD, we first ignore the propagation delay. Let $\mu$ be the source, $\nu$ the destination, $t(\mu, \nu)$ the MMD from $\mu$ to $\nu$, and $d(\mu, \nu)$ the distance between $\mu$ and $\nu$. We show that, a.s.

$$\lim_{d(\mu,\nu)\to\infty} \frac{t(\mu,\nu)}{d(\mu,\nu)} \begin{cases} = 0, & \text{if instantaneously connected;} \\ > 0, & \text{if intermittently connected.} \end{cases}$$

When the secondary network is instantaneously connected, a much stronger statement is actually shown, that is,

$$\lim_{d(\mu,\nu)\to\infty} \frac{t(\mu,\nu)}{g(d(\mu,\nu))} = 0 \text{ a.s.,}$$

where $g(d(\mu,\nu))$ is any monotonically increasing function of $d(\mu,\nu)$ satisfying $\lim_{d(\mu,\nu)\to\infty} g(d(\mu,\nu)) = \infty$. It implies that the MMD $t(\mu,\nu)$ is asymptotically independent of the distance $d(\mu,\nu)$ as $d(\mu,\nu) \to \infty$. Thus when the propagation delay is negligible, an instantaneously connected CR network behaves almost the same as a homogeneous ad hoc network in the sense that the waiting time for spectrum opportunities does not affect the scaling law of the MMD with respect to the distance. On the other hand, if a CR network is intermittently connected, the waiting time for the spectrum opportunities accumulates linearly with the source-destination distance, resulting in a fundamental difference in the MMD as compared to a homogeneous network.

The above scaling law may be illustrated by an analogy of traveling from a place $\mu$ to another place $\nu$, where the waiting time for the spectrum opportunities is likened to the waiting time for traffic lights. Suppose that we can move fast enough such that the driving time on the road is negligible. When the secondary network is instantaneously connected, there exists a.s. an infinite connected component consisting of communication links which can be considered a highway without traffic lights between $\mu$ and $\nu$. Given that both $\mu$ and $\nu$ are within a finite distance to the highway (independent of the distance between $\mu$ and $\nu$), the travel time from $\mu$ to $\nu$, which is the waiting time for traffic lights before entering the highway and after leaving the highway, is independent of the distance between $\mu$ and $\nu$. When the secondary network is intermittently connected, such a highway between $\mu$ and $\nu$ can not be found. Then we have to use local paths and wait for traffic lights from time to time, leading to the linear scaling of the travel time with respect to the distance between $\mu$ and $\nu$ even when the driving time is ignored.

We also study the impact of the propagation delay on the MMD. When the propagation delay $\tau$ is nonnegligible, we show that the MMD scales linearly with the source-destination distance in both instantaneously connected and intermittently connected regimes, but with different rates.

In particular, the limiting behavior of the rate as $\tau \to 0$ is distinct in the two regimes, *i.e.,* a.s.

$$\lim_{\tau \to 0} \lim_{d(\mu,\nu) \to \infty} \frac{t(\mu,\nu)}{d(\mu,\nu)} \begin{cases} = 0, & \text{if instantaneously connected;} \\ > 0, & \text{if intermittently connected.} \end{cases}$$

It indicates that when the propagation delay is sufficiently small, the scaling rate of the MMD for an instantaneously connected network is much smaller than the one for an intermittently connected network.

### B. Related Work

As a fundamental indicator of the feasibility and efficiency of large-scale wireless networks, the scaling law has received increasing interest in the research community since the seminal work of Gupta and Kumar [3]. The capacity scaling law of CR networks has been analyzed in [4, 5, 6]. In [4], the authors also derive the capacity-delay tradeoff for a specific routing and scheduling algorithm which is shown to achieve the optimal one for homogeneous networks. To our best knowledge, the scaling law of the MMD with respect to the source-destination distance in a CR network has not been characterized in the literature.

The scaling law of the multihop delay in homogeneous ad hoc networks has been well studied (see [7, 8, 9, 10, 11, 12, 13, 14, 15, 16, 17] and references therein). The multihop delay for a specific routing algorithm is analyzed in [7, 8, 9], and the capacity-delay tradeoff is established under a given network and mobility model in [10, 11, 12]. Theoretical bounds on the information propagation speed defined as the ratio of the travel distance to the multihop delay are derived for a static network in [13, 14] or for a mobile network in [15]. Based on continuum percolation theory, the scaling law of the multihop delay with respect to the source-destination distance is established in [16, 17]. In particular, Kong and Yeh considered in [17] homogeneous ad hoc networks with dynamic on-off links and showed that the scaling of the MMD behaves distinctly in two regimes, depending on whether the network is percolated. In this paper, we use techniques in continnum percolation that are similar to those used in [17]. A major difference is that the states of the links in the secondary network considered in this paper are correlated instead of independent, which complicates the analysis of multihop delay.

There are also a number of results on intermittently connected networks (see, for example, [18, 19, 20, 21]), where the intermittent connectivity is caused by node mobility or duty cycling, instead of spatial and temporal dynamics of spectrum opportunities. The problem and the technical approach are generally different.

*C. Organization and Notation*

The rest of the paper is organized as follows. In Sec. II, the Poisson CR network model is specified. Based on a close examination of the conditions for the existence of a communication link in CR networks, the connectivity of the CR network is analytically characterized in Sec. III. Sec. IV presents the results about the scaling behavior of the MMD with respect to the source-destination distance. Simulation results are provided in Sec. V to illustrate the analytical results proven in Sec. IV. Sec. VI concludes the paper.

Throughout the paper, we use capital letters for parameters of primary users and lowercase letters for secondary users.

## II. NETWORK MODEL

We consider a Poisson distributed secondary network overlaid with a Poisson distributed primary network in an infinite two dimensional Euclidean space[1]. The primary network adopts a synchronized slotted structure with a slot length $T_S$. The realizations of active primary transmitters vary from slot to slot and are assumed to be i.i.d. across slots[2]. Thus $T_S$ can be considered as the time constant of the spectrum opportunities which are determined by the transmitting and receiving activities of the primary users. Without loss of generality, we set $T_S = 1$.

At the beginning of each slot, the primary transmitters are distributed according to a two-dimensional Poisson point process $X_{PT}$ with density $\lambda_{PT}$. Primary receivers are uniformly distributed within the transmission range $R_p$ of their corresponding transmitters. Here we have assumed that all the primary transmitters use the same transmission power and the transmitted signals undergo an isotropic path loss. Based on the displacement theorem [22, Chapter 5], it is easy to see that the primary receivers form another two-dimensional Poisson point process $X_{PR}$ with density $\lambda_{PT}$, which is correlated with $X_{PT}$.

The secondary users are distributed according to a two-dimensional Poisson point process $X_S$ with density $\lambda_S$, which is independent of $X_{PT}$ and $X_{PR}$. The locations of the secondary users are static over time, and they have a uniform transmission range $r_p$.

---

[1]This infinite network model is equivalent in distribution to the limit of a sequence of finite networks with a fixed density as the area of the network increases to infinity, *i.e.,* the so-called *extended network*.

[2]The different realizations of active primary transmitters in different slots can be caused by the mobility of these users or changes in the traffic pattern or both.

## III. CONNECTIVITY

In this section, we examine the connectivity of the secondary network by analytically characterizing the partition of the $(\lambda_S, \lambda_{PT})$ plane illustrated in Fig. 1.

### A. Topological Link vs. Communication Link

Topological links in the secondary network are independent of the primary network. A topological link exists between any two secondary users that are within each other's transmission range. In contrast, the existence of a communication link between two secondary users depends not only on the distance between them but also on the availability of the communication channel, *i.e.,* the presence of a spectrum opportunity offered by the primary network. As a result, even in a static secondary network, communication links are time-varying due to the temporal dynamics of spectrum opportunities. The presence of a spectrum opportunity is determined by the transmitting and receiving activities of the primary network as described below.

We consider the disk signal propagation and interference model as illustrated in Fig. 2. There exists an opportunity from $\mu$, the secondary transmitter, to $\nu$, the secondary receiver, if the transmission from $\mu$ does not interfere with *primary receivers* in the solid circle, and the reception at $\nu$ is not affected by *primary transmitters* in the dashed circle [23]. Referred to as the interference range of the secondary users, the radius $r_I$ of the solid circle at $\mu$ depends on the transmission power of $\mu$ and the interference tolerance of the primary receivers, whereas the radius $R_I$ of the dashed circle (the interference range of the primary users) depends on the transmission power of the primary users and the interference tolerance of $\nu$.

It is clear from the above discussion that spectrum opportunities are *asymmetric*. Specifically, a channel that is an opportunity when $\mu$ is the transmitter and $\nu$ the receiver may not be an opportunity when $\nu$ is the transmitter and $\mu$ the receiver. Since unidirectional links are difficult to utilize, especially for applications with guaranteed delivery that require acknowledgements, we only consider bidirectional links in the secondary network when we define connectivity.

### B. Connectivity and the Finiteness of MMD

As stated in Sec. I, the connectivity of the secondary network is defined by the finiteness of the MDD between two randomly chosen secondary users. In this section, we show that while the transmissions between two secondary users can only be carried by communication links, the finiteness of the MMD depends solely on the topological connectivity of the secondary network when the temporal dynamics of the primary traffic is sufficiently rich.

Consider an undirected random graph $\mathcal{G}_S(\lambda_S)$ consisting of all the secondary users and the topological links. Notice that $\mathcal{G}_S(\lambda_S)$ depends only on the Poisson point process $X_S$ of the secondary network. Under the i.i.d. model of the temporal dynamics of the primary traffic, we show in Theorem 1 below that a necessary and sufficient condition for the a.s. finiteness of the MMD in the secondary network is the connectivity of $\mathcal{G}_S(\lambda_S)$ in the percolation sense.

*Theorem 1:* Let $t(\mu, \nu)$ denote the MMD between two randomly chosen secondary users $\mu$ and $\nu$. Then with a positive probability, $t(\mu, \nu) < \infty$ a.s. if and only if $\lambda_S > \lambda_c$ where $\lambda_c$ is the critical density of homogeneous ad hoc networks.

*Proof:* It follows from the classic result on homogeneous networks [24, Chapter 3] that there exists an infinite connected component in $\mathcal{G}_S(\lambda_S)$ if and only if $\lambda_S > \lambda_c$, where $\lambda_c$ is the critical density of homogeneous networks.

If $\lambda_S \leq \lambda_c$, then there exist only finite connected components in $\mathcal{G}_S(\lambda_S)$; there is no topological path between $\mu$ and $\nu$ a.s., *i.e.,* $t(\mu, \nu) = \infty$ a.s. On the other hand, if $\lambda_S > \lambda_c$, then with a positive probability $\mu$ and $\nu$ belong to the infinite topologically connected component[3]. In other words, there exists a topological path $L$ with finite number of hops between $\mu$ and $\nu$. Let $t^L(\mu, \nu)$ denote the multihop delay from $\mu$ to $\nu$ along the path $L$. Next we prove the a.s. finiteness of $t^L(\mu, \nu)$ by showing the following lemma about the single-hop delay.

*Lemma 1:* Let $t_s(w_1, w_2)$ denote the single-hop delay from $w_1$ to $w_2$, where $w_1$ and $w_2$ are connected via a topological link. Then we have that $t_s(w_1, w_2) < \infty$ a.s.

*Proof of Lemma 1:* see Appendix A. ∎

Since $t^L(\mu, \nu)$ is a finite sum of single-hop delays, we have that $t^L(\mu, \nu) < \infty$ a.s. Thus, $t(\mu, \nu) \leq t^L(\mu, \nu) < \infty$ a.s. ∎

Theorem 1 shows that under the i.i.d. model of the temporal dynamics of the primary traffic, the connectivity of the secondary network defined by the finiteness of the MMD is equivalent to the *topological* connectivity of $\mathcal{G}_S(\lambda_S)$ which is independent of the primary network. In other words, no matter how heavy the primary traffic is, the MMD between two secondary users in the infinite topologically connected component of $\mathcal{G}_S(\lambda_S)$ is finite a.s.

---

[3]It is shown in [24, Theorem 3.6] that when $\lambda_S > \lambda_c$, there exists a unique infinite connected component in $\mathcal{G}_S(\lambda_S)$ a.s.

We point out that the i.i.d. model of the temporal dynamics of the primary traffic is not necessary for Theorem 1 to hold. This i.i.d. model can be considered as one end of the spectrum on the richness of the temporal dynamics of the primary traffic. The other end of the spectrum is given by a static set of primary transmitters and receivers. In this case, the finiteness of MMD can only be achieved through instantaneous connectivity using only communication links. It is an interesting future direction to obtain necessary conditions on the temporary dynamics of the primary traffic that ensures the equivalence between the finiteness of MMD and the topological connectivity of $\mathcal{G}_S(\lambda_S)$. From the proof of Theorem 1 we can see that this equivalence holds whenever the temporary dynamics of the primary traffic makes the single-hop delay have a proper distribution.

### C. Instantaneous Connectivity vs. Intermittent Connectivity

In a primary slot $t$, we can obtain an undirected random graph $\mathcal{G}_H(\lambda_S, \lambda_{PT}, t)$ consisting of all the secondary users and the communication links which represents the instantaneous connectivity of the secondary network in this slot. As illustrated in Fig. 3, this graph $\mathcal{G}_H(\lambda_S, \lambda_{PT}, t)$ is determined by the three Poisson point processes in slot $t$: $X_S$, $X_{PT}$, and $X_{PR}$, where $X_{PT}$ and $X_{PR}$ are correlated.

We define the instantaneous connectivity of the secondary network as the a.s. existence of an infinite connected component in $\mathcal{G}_H(\lambda_S, \lambda_{PT}, t)$ for all $t$. Given the transmission power and the interference tolerance of both the primary and the secondary users ($i.e.,\ R_p,\ R_I,\ r_p$, and $r_I$ are fixed), the instantaneous connectivity region $\mathcal{C}$ is defined as

$$\mathcal{C} \triangleq \{(\lambda_S,\ \lambda_{PT}):\ \mathcal{G}_H(\lambda_S, \lambda_{PT}, t) \text{ is connected for all } t\}.$$

A detailed analytical characterization of $\mathcal{C}$ is given in [25]. Let $\theta(\lambda_S, \lambda_{PT})$ denote the probability that an arbitrary secondary user belongs to an infinite connected component[4] in $\mathcal{G}_H(\lambda_S, \lambda_{PT}, t)$, then we have that

$$\theta(\lambda_S, \lambda_{PT}) \begin{cases} > 0, & \text{if } (\lambda_S, \lambda_{PT}) \in \mathcal{C}; \\ = 0, & \text{otherwise.} \end{cases} \tag{1}$$

Referred to as the critical density of the secondary users, $\lambda_S^*$ is the infimum density of the secondary users that ensures instantaneous connectivity under a positive density of primary

---

[4]Since the distribution of the primary network is i.i.d. across slots, it is easy to see that this probability $\theta$ is time-invariant.

transmitters:

$$\lambda_S^* \triangleq \inf\{\lambda_S : \exists \lambda_{PT} > 0 \text{ such that } \mathcal{G}_H(\lambda_S, \lambda_{PT}, t) \text{ is connected for all } t\}.$$

It is shown in [25] that $\lambda_S^*$ equals the critical density $\lambda_c$ of a *homogeneous* ad hoc network.

$\mathcal{G}_H(\lambda_S, \lambda_{PT}, t)$ can also be obtained by removing topological links that do not see the opportunities in slot $t$ from the random graph $\mathcal{G}_S(\lambda_S)$. Thus, even if the secondary network is connected (*i.e.,* $\mathcal{G}_S(\lambda_S)$ has an infinite connected component), it may not be instantaneously connected. Specifically, the infinite connected component in $\mathcal{G}_S(\lambda_S)$ may break into infinite number of finite connected components in $\mathcal{G}_H(\lambda_S, \lambda_{PT}, t)$ due to scarcity of spectrum opportunities. In this case, we define the intermittent connectivity region $\mathcal{C}_I$ as

$$\mathcal{C}_I \triangleq \{(\lambda_S, \ \lambda_{PT}) : \ \lambda_S > \lambda_c \text{ and } \mathcal{G}_H(\lambda_S, \lambda_{PT}, t) \text{ is disconnected for all } t\}.$$

## IV. MULTIHOP DELAY

In this section, we analytically characterize the scaling behavior of the MMD with respect to the source-destination distance. Let $C(\mathcal{G}_S(\lambda_S))$ be the infinite connected component in $\mathcal{G}_S(\lambda_S)$ when $\lambda_S > \lambda_c$, *i.e.,* the secondary network is connected. We seek to establish the scaling law of the MMD between two arbitrary users in $C(\mathcal{G}_S(\lambda_S))$ with respect to the distance between them. As shown in the following two subsections which consider the two cases when the propagation delay $\tau = 0$ and $\tau > 0$, whether the secondary network is instantaneously connected or intermittently connected determines the scaling behavior of the MMD.

### A. Negligible Propagation Delay

When the propagation delay $\tau = 0$, once a user has received the message, it can spread the message instantaneously throughout the connected component formed by communication links which contains it. Thus, if the secondary network is instantaneously connected, the source can route its message via the infinite connected component such that the message can make a huge step towards the destination within a single primary slot. As we will see, this huge step in the infinite connected component leads to the asymptotic independence of the multihop delay on the source-destination distance. On the other hand, if the secondary network is intermittently connected, the message can move forward only a limited step within each primary slot, which results in the linear scaling of the MMD. A mathematical statement about the scaling is given in the following theorem.

*Theorem 2:* Assume that $\tau = 0$. For any two secondary users $\mu$, $\nu \in C(\mathcal{G}_S(\lambda_S))$, where $C(\mathcal{G}_S(\lambda_S))$ is the infinite connected component of $\mathcal{G}_S(\lambda_S)$, let $t(\mu, \nu)$ denote the MMD from $\mu$ to $\nu$ and $d(\mu, \nu)$ the distance between $\mu$ and $\nu$; then

T2.1  if $(\lambda_S, \lambda_{PT}) \in \mathcal{C}$,

$$\lim_{d(\mu,\nu) \to \infty} \frac{t(\mu, \nu)}{g(d(\mu, \nu))} = 0 \text{ a.s.,}$$

where $g(d)$ is any monotonically increasing function of $d$ with $\lim_{d \to \infty} g(d) = \infty$;

T2.2  if $(\lambda_S, \lambda_{PT}) \in \mathcal{C}_I$, $\exists \; 0 < \beta < \infty$ such that

$$\lim_{d(\mu,\nu) \to \infty} \frac{t(\mu, \nu)}{d(\mu, \nu)} = \beta \text{ a.s.,} \tag{2}$$

where the value of $\beta$ depends on $(\lambda_S, \lambda_{PT})$.

To simplify the notation, let the minimum path denote the path from the source to the destination with the MMD. Notice that the minimum path depends on both the topology of the secondary network and the realizations of the primary network over time. In other words, the minimum path for a realization of the primary network may not be the minimum path for another realization of the primary network. It is, thus, intractable to directly study the minimum path between the source and the destination. Instead, we analyze the multihop delay along a constructed path to provide an upper bound on the MMD for an instantaneously connected network, and derive a lower bound on the MMD for an intermittently connected network by considering a.s. finiteness of the connected components formed by communication links. In the proof, we borrow techniques and theories from continuum percolation and ergodicity, including the discretization technique, the FKG inequality, and the Subadditive Ergodic Theorem [26].

*Proof of T2.1:*  We use the infinite connected component consisting of communication links[5] in $\mathcal{G}_H(\lambda_S, \lambda_{PT}, t_0)$ during some primary slot $t_0$ to construct a path $L_C$ from $\mu$ to $\nu$ such that the multihop delay along this path is independent of the distance $d(\mu, \nu)$ (see Fig. 4 for an illustration). Then we analyze the multihop delay $t^C(\mu, \nu)$ along $L_C$.

Assume that $\mu$ starts trying to send the message at time $t = 0$. Let $C(t)$ be the infinite connected component in $\mathcal{G}_H(\lambda_S, \lambda_{PT}, t)$, and $t_0$ the first primary slot such that $\mu \in C(t_0)$. Based on (1), we know that the probability $\theta(\lambda_S, \lambda_{PT})$ that $\mu \in C(t)$ for each $t$ is strictly positive. It follows from the i.i.d. distribution of the primary network across slots that $t_0$ is finite a.s. Given $C(t_0)$, we define user $w_\nu$ as the user in $C(t_0)$ which is closest to $\nu$, *i.e.*,

$$w_\nu \overset{\Delta}{=} \underset{w_i \in C(t_0)}{\arg\min} \ d(w_i, \nu).$$

---

[5]It is shown in [25] that there exists either zero or one infinite connected component in $\mathcal{G}_H(\lambda_S, \lambda_{PT}, t)$ a.s. for any given $t$.

Notice that if $\nu \in C(t_0)$, then $w_\nu = \nu$.

As illustrated in Fig. 4, the constructed path $L_C$ passes through $w_\nu$, then the multihop delay $t^C(\mu, \nu)$ along the path $L_C$ can be expressed as:

$$t^C(\mu, \nu) = t_0 + t(\mu, w_\nu) + t(w_\nu, \nu) = t_0 + t(w_\nu, \nu),$$

where $t(w_\nu, \nu)$ is the MMD from $w_\nu$ to $\nu$. In the last step, we have used $t(\mu, w_\nu) = 0$, since $\mu$, $w_\nu \in C(t_0)$ and $\tau = 0$. Next we show the following lemma.

*Lemma 2:* $t(w_\nu, \nu)$ is finite a.s.

*Proof of Lemma 2:* We first show that $d(w_\nu, \nu) < \infty$ a.s. by using the ergodicity of the CR network model, and then obtain an upper bound on the multihop delay along the shortest path[6] $L(w_\nu, \nu)$ from $w_\nu$ to $\nu$. Since $t(w_\nu, \nu) \leq t^L(w_\nu, \nu)$ where $t^L(w_\nu, \nu)$ is the multihop delay along $L(w_\nu, \nu)$, the a.s. finiteness of $t(w_\nu, \nu)$ follows from that of the upper bound on $t^L(w_\nu, \nu)$. The proof here is inspired by the proof of Lemma 9 in [17], but with a much simpler proof of $d(w_\nu, \nu) < \infty$. For details, see Appendix B. ∎

It is easy to see that $t_0$ and $t(w_\nu, \nu)$ are independent of $d(\mu, \nu)$. Then we conclude that

$$\lim_{d(\mu, \nu) \to \infty} \frac{t^C(\mu, \nu)}{g(d(\mu, \nu))} = 0 \text{ a.s.},$$

and T2.1 follows immediately from the fact that $t(\mu, \nu) \leq t^C(\mu, \nu)$. ∎

*Proof of T2.2:* Based on the scaling argument [24, Chapter 2], we set the transmission range $r_p$ of the secondary users to 1 without loss of generality. Take $\mu$ as the origin, and the line connecting $\mu$ and $\nu$ as the x-axis. Define an auxiliary node $\tilde{w}_i$ in $C(\mathcal{G}_S(\lambda_S))$ for every integer $i$:

$$\tilde{w}_i \overset{\Delta}{=} \underset{w \in C(\mathcal{G}_S(\lambda_S))}{\arg\min} \ d(w, (i, 0)).$$

---

[6] The shortest path is the path from the source to the destination with the minimum number of hops. Notice that the shortest path is not necessarily the minimum path, since the probability of having an opportunity is a function of the hop length and a longer hop usually results in more waiting time.

Obviously, $\tilde{w}_0 = \mu$. Let $n$ be the closest integer to $\nu$, then

$$\frac{t(\tilde{w}_0, \tilde{w}_n) - t(\tilde{w}_n, \nu)}{n + 1} \leq \frac{t(\mu, \nu)}{d(\mu, \nu)} \leq \frac{t(\tilde{w}_0, \tilde{w}_n) + t(\tilde{w}_n, \nu)}{n - 1}.$$

If $\tilde{w}_n = \nu$, then $t(\tilde{w}_n, \nu) = 0$; if $\tilde{w}_n \neq \nu$, then $t(\tilde{w}_n, \nu)$ is at most the single-hop delay because $d(\tilde{w}_n, \nu) \leq d(\tilde{w}_n, (n, 0)) + d(\nu, (n, 0)) \leq 2d(\nu, (n, 0)) \leq 1$.

Let $t_{m,n} = t(\tilde{w}_m, \tilde{w}_n)$ for any two integers $m$, $n$. Then to show T2.2, it suffices to show that

$$\lim_{n \to \infty} \frac{t_{0,n}}{n} = \beta > 0 \text{ a.s.} \tag{3}$$

The proof of (3) is divided into the following two lemmas.

*Lemma 3:* $\beta \overset{\triangle}{=} \lim\limits_{n \to \infty} \frac{t_{0,n}}{n}$ exists a.s.

*Lemma 4:* $0 < \beta = \lim\limits_{n \to \infty} \frac{t_{0,n}}{n} < \infty$.

The proof of Lemma 3 is based on the Subadditive Ergodic Theorem [26, Theorem 1.10], and the proof of Lemma 4 is based on the fact about the diameter[7] of the finite connected components formed by communication links in an intermittently connected network. For details, see Appendix C and Appendix D. ∎

### B. Nonnegligible Propagation Delay

When the propagation delay $\tau > 0$, it takes at least $\tau$ for the message to traverse a distance $r_p$, which imposes a lower bound $\tau/r_p$ on the ratio of the MMD to the source-destination distance. This implies that the MMD scales at least linearly with the source-destination distance.

The positive propagation delay $\tau$ also imposes an upper bound $T_S/\tau$ on the maximum number of hops that the message can traverse in a primary slot $T_S$. For an instantaneously connected network, this upper bound can be actually attained in the infinite connected component consisting of communication links. But for an intermittently connected network, this upper bound may probably not be attained due to the limited diameter of the finite connected components formed by communication links, especially when the propagation delay $\tau$ is small. Specifically, there may not exist a connected component which has a path with $T_S/\tau$ hops. Thus, although the scaling order is always linear, it can be expected that the scaling rate for an instantaneously connected network is much smaller than the one for an intermittently connected network. The following theorem summarizes the above observations in a rigorous form.

---

[7]The diameter of a connected component $C$ is defined as $\max\limits_{\mu, \nu \in C} d(\mu, \nu)$.

*Theorem 3:* Assume that $\tau > 0$. For any two secondary users $\mu$, $\nu \in C(\mathcal{G}_S(\lambda_S))$, where $C(\mathcal{G}_S(\lambda_S))$ is the infinite connected component of $\mathcal{G}_S(\lambda_S)$, let $t^\tau(\mu, \nu)$ denote the MMD from $\mu$ to $\nu$ and $d(\mu, \nu)$ the distance between $\mu$ and $\nu$; then $\exists \; \gamma = \gamma(\tau) > 0$ such that

$$\lim_{d(\mu,\nu)\to\infty} \frac{t^\tau(\mu, \nu)}{d(\mu, \nu)} = \gamma \geq \frac{\tau}{r_p} \text{ a.s.}. \tag{4}$$

Furthermore, if $(\lambda_S, \lambda_{PT}) \in \mathcal{C}$,

$$\lim_{\tau\to 0} \lim_{d(\mu,\nu)\to\infty} \frac{t^\tau(\mu, \nu)}{d(\mu, \nu)} = 0 \text{ a.s.}; \tag{5}$$

if $(\lambda_S, \lambda_{PT}) \in \mathcal{C}_I$,

$$\lim_{\tau\to 0} \lim_{d(\mu,\nu)\to\infty} \frac{t^\tau(\mu, \nu)}{d(\mu, \nu)} \geq \beta > 0 \text{ a.s.}, \tag{6}$$

where $\beta = \lim\limits_{d(\mu,\nu)\to\infty} \frac{t(\mu,\nu)}{d(\mu,\nu)}$ is defined in (2).

*Proof Sketch:* The equality in (4) is based on the Subadditive Ergodic Theorem [26], while the inequality in (4) is a direct consequence of the simple observation above. The basic idea behind establishing (5) is to consider the multihop delay along the path constructed in the proof of T2.1. Eqn. (6) follows immediately from the fact that $t^\tau(\mu, \nu) \geq t(\mu, \nu)$, where $t(\mu, \nu)$ is the MMD when $\tau = 0$. For details, see Appendix E. ∎

## V. Simulation Results

In this section, we present several simulation results. The density $\lambda_S$ of the simulated secondary network is larger than the critical density $\lambda_c$. Thus, the secondary network is either instantaneously connected or intermittently connected, depending on the density $\lambda_{PT}$ of the primary transmitters. Without loss of generality, we assume that the source is located at the origin. Each node in the network is a potential destination. This allows us to simulate different realizations of the source-destination pair using one Monte Carlo run.

We obtain the MMD by considering the flooding scheme. Specifically, every user which has received the message (including the source) will transmit the message to its neighbors within its transmission range when it experiences a bidirectional spectrum opportunity to any of its neighbors. The transmission attempts will not stop until all its neighbors receive the message. The time that a user first receives the message during the flooding is considered as the MMD from the source to this user. It is easy to see that simulating this flooding scheme gives us the MMD when there is no contention between the secondary users' transmissions.

Fig. 5 shows the MMD-to-distance ratio as a function of the source-destination distance when the propagation delay $\tau$ is zero, where each dot represents a realization of the destination. We can see that if the secondary network is instantaneously connected (Fig. 5-(a)), the ratio decreases very fast as the distance increases, and it can be expected that the ratio will eventually tend to zero. On the other hand, if the secondary network is intermittently connected (Fig. 5-(b)), the decreasing rate of the ratio levels off as the distance increases, and the ratio will gradually approach a positive constant. Note that in Fig. 5-(a), the MMD-to-distance ratios of different realizations of the destination are grouped into several continuous curves, each associated with a fixed MMD. Specifically, since the message is mainly delivered via the infinite connected component consisting of communication links when the secondary network is instantaneously connected, the secondary users are actually grouped according to the first time that they are in an infinite connected component. From Fig. 5-(a) we can see that due to the temporal dynamics of spectrum opportunities, every node will be part of an infinite connected component within a few number of primary slots.

In Fig. 5, we compare the MMD-to-distance ratio in an instantaneously connected network and in an intermittently connected network when the propagation delay $\tau$ is nonzero but small. The two red dashed lines in Fig. 5(c)(d) denote the lower bound $\tau/r_p$ on the ratio imposed by the propagation delay. Although the ratio for the instantaneously connected network does not go to zero due to the nonnegligible propagation delay, it is $10$ times smaller than the ratio for the intermittently connected network.

## VI. Conclusion and Discussion

We have studied the connectivity and multihop delay of ad hoc cognitive radio networks. The impact of connectivity on the multihop delay has been examined by establishing the scaling behavior of the minimum multihop delay with respect to the source-destination distance. Specifically, depending on whether the cognitive radio network is instantaneously connected or intermittently connected, the scaling of the minimum multihop delay behaves distinctly, in terms of either the scaling order when the propagation delay is negligible or the scaling rate when the propagation delay is nonnegligible. This result on scaling is independent of the random positions of the source and the destination, and it only depends on the network parameters (e.g., the density of the secondary users and the traffic load of the primary network). In establishing these results, we have used theories and techniques from continuum percolation and ergodicity including the concept of critical density, the FKG inequality, the discretization technique, and the Subadditive Ergodic Theorem.

In the above analysis, we have assumed a disk signal propagation model which only incorporates the path-loss. If we take into account fading, then the condition for the existence of a topological link between two secondary users should be changed into the received SNR at each user. Since a fixed transmission range does not hold here, this leads to a random connection model (RCM) [24, Chapter 1] where, for any two users in the network, there exists a link connecting them with some probability (maybe zero) depending on the distance between them. Considering that the RCM shares several basic properties (e.g., the ergodicity and the existence of the critical density) with the Boolean model used in this paper, we expect that the results established here can be extended to the RCM, although the derivations may become more complicated.

We have also assumed that the interference region can be represented by a circle with a fixed radius. It is possible that interference aggregated from multiple interferers outside the interference region cause an outage at the receiver. By choosing a conservative interference range, however, this possibility is negligible [27].

## Appendix A: Proof of Lemma 1

Assume that $\tau \le T_S = 1$ such that the spectrum opportunity lasts long enough to ensure the success of the transmission, and $w_1$ intends to transmit the message at time 0. It follows from Sec. III-A that the single-hop delay $t_s(w_1, w_2)$ is the waiting time $t_{sw}(w_1, w_2)$ for the presence of the first bidirectional opportunity plus the propagation delay $\tau$, $i.e.$,

$$t_s(w_1, w_2) = t_{sw}(w_1, w_2) + \tau = \argmin_{n \in \{0, 1, 2, \dots\}} \{ \mathbb{I}_{(w_1, w_2)}(n) = 1 \} + \tau,$$

where $\mathbb{I}_{(w_1, w_2)}(n)$ be an indicator such that $\mathbb{I}_{(w_1, w_2)}(n) = 1$ if a bidirectional opportunity exists between $w_1$ and $w_2$ during the $n$th primary slot, and $\mathbb{I}_{(w_1, w_2)}(t) = 0$ otherwise.

Due to the i.i.d. distribution of the primary network across slots, $t_{sw}(w_1, w_2)$ is a geometric random variable with parameter $p_0$, where $p_0$ is the probability of having a bidirectional opportunity between $w_1$ and $w_2$ at any given time and is always strictly positive [28, Appendix A]. Thus, $t_{sw}(w_1, w_2) < \infty$ a.s., and $t_s(\mu, \nu) < \infty$ a.s.

## Appendix B: Proof of Lemma 2

We first establish that $d(w_\nu, \nu) < \infty$. Since $d(w_\nu, \nu) \leq d(w_\nu, (0,0)) + d(\nu, (0,0))$ and $d(\nu, (0,0)) < \infty$ a.s., it follows that $d(w_\nu, \nu) < \infty$ a.s. if $d(w_\nu, (0,0)) < \infty$ a.s. Consider the following three events:

$$
\begin{aligned}
E &= \{d(w_\nu, (0,0)) < \infty\}, \\
E_r &= \{\ w_\nu \in C(t_0) \text{ such that } d(w_\nu, (0,0)) \leq r\}, \\
E_{r1} &= \{\ w_\nu \in \mathcal{G}_S(\lambda_S) \text{ such that } d(w_\nu, (0,0)) \leq r\}.
\end{aligned}
$$

Then we have that for a fixed $r > 0$,

$$
\Pr\{E\} \geq \Pr\{E_r\} \geq \Pr\{E_{r1}\}\theta(\lambda_S, \lambda_{PT}) = [1 - \exp(-\lambda_S \pi r^2)]\theta(\lambda_S, \lambda_{PT}) > 0,
$$

where $\theta(\lambda_S, \lambda_{PT})$ is defined in (1) and is strictly positive since $(\lambda_S, \lambda_{PT}) \in \mathcal{C}$. It is easy to see that the event $E$ is invariant of the shift transformations[8]. Thus, based on the ergodicity[9] of the CR network model [25], we conclude that $\Pr\{E\} = 1$, *i.e.,* $d(w_\nu, (0,0)) < \infty$ a.s.

Next we show that the multihop delay $t^L(w_\nu, \nu)$ along the shortest path $L(w_\nu, \nu)$ between $w_\nu$ and $\nu$ is finite a.s. We do this in multiple steps. Without loss of generality, we set $r_p$ to be 1.

First we show that there exists a sequence of topologically connected nodes intersecting squares containing $w_\nu$ and $\nu$ with positive probability. Second, we show that a closed circuit of connected users exists in each square. Third, we show that a finite hop-length path from $w_\nu$ to $\nu$ exists within a square. Finally, we show that this implies a finite multihop-delay path from $w_\nu$ to $\nu$.

Step 1: We construct a sequence of concentric squares with increasing side lengths as illustrated in Fig. 6. Specifically, all the squares are centered at the midpoint of the segment joining $w_\nu$ and $\nu$, and the side length of the $j$-th ($j \geq 0$) square $S_j$ is $3^j d$. Let $A_j$ ($j \geq 1$) denote the square annulus inside $S_j$ and outside $S_{j-1}$, and let $E_j^u$ be the event that there exists a left-to-right crossing[10] in the upper horizontal rectangle of $A_j$ with side length $3^j d \times 3^{j-1} d$. Similarly, define $E_j^b$, $E_j^l$, and $E_j^r$ as the events that the bottom, left, and right rectangles of $A_j$ are crossed from left to right or from top to bottom. By symmetry, we know that $\Pr(E_j^u) = \Pr(E_j^b) = \Pr(E_j^l) = \Pr(E_j^r)$.

---

[8]For a random model in a Euclidean space $\mathbb{R}^d$ with a probability space $(\Omega, \mathcal{F}, \mu)$, the shift transformation $S_x$ is to shift the realization $\omega \in \Omega$ by $x \in \mathbb{R}^d$.

[9]A random model under a probability space $(\Omega, \mathcal{F}, \mu)$ is said to be ergodic if there exists a transformation group $\{S_x : x \in \mathbb{R}^d \text{ or } \mathbb{Z}^d\}$ that acts ergodically on $(\Omega, \mathcal{F}, \mu)$. A transformation group $\{S_x : x \in \mathbb{R}^d \text{ or } \mathbb{Z}^d\}$ is said to act ergodically if the $\sigma$-algebra of events invariant under the whole group is trivial, *i.e.,* any invariant event has measure either zero or one. For an ergodic random model $(\Omega, \mathcal{F}, \mu)$, if an event $E \in \mathcal{F}$ invariant under the whole transformation group $\{S_x : x \in \mathbb{R}^d \text{ or } \mathbb{Z}^d\}$ occurs with a positive probability, *i.e.,* $\mu(E) > 0$, then it occurs a.s., *i.e.,* $\mu(E) = 1$.

[10]A left-to-right crossing exists in a rectangle $R = [x_1, x_2] \times [y_1, y_2]$ if and only if there exists a sequence of nodes $\mu_i$ ($1 \leq i \leq n$) in $\mathcal{G}(\lambda_S)$ such that (i) $\mu_i \in R$ for all $i$; (ii) $d(\mu_{i+1}, \mu_i) \leq 1$ for all $1 \leq i < n$; (iii) $|x(\mu_1) - x_1| \leq \frac{1}{2}$ and $|x(\mu_n) - x_2| \leq \frac{1}{2}$, where $x(\mu_i)$ is the x-coordinate of $\mu_i$. The top-to-bottom crossing can be defined analogously. Note that the sequence of nodes constituting the crossing represent a topologically connected path.

Since $(\lambda_S, \lambda_{PT}) \in \mathcal{C}$, it follows that $\lambda_S > \lambda_c$ where $\lambda_c$ is the critical density for a homogeneous network. By using Corollary 4.1 in [24], we have that $\lim_{d \to \infty} \Pr\{E_1^u\} = 1$. Then for a given $\delta$, $0 < \delta < 1$, we choose

$$d = d_\delta \triangleq \max\{\inf\{d' : \Pr\{E_1^u\} \geq \delta \text{ if } d \geq d'\}, d(w_\nu, \nu)\}.$$

We then have that $\Pr\{E_j^u\} \geq \delta > 0$ for all $j \geq 1$.

Step 2: Let $E_j$ $(j \geq 1)$ be the event that there exists a closed circuit of connected users in $\mathcal{G}(\lambda_S)$ within $A_j$. If $E_j^u$, $E_j^b$, $E_j^l$, and $E_j^r$ all occur, then $E_j$ occurs (see Fig. 6). Since $E_j^u$, $E_j^b$, $E_j^l$, and $E_j^r$ are all increasing events[11], it follows from the FKG inequality [24, Theorem 2.2] that

$$\Pr\{E_j\} \geq \Pr\{E_j^u \cap E_j^b \cap E_j^l \cap E_j^r\} \geq \Pr\{E_j^u\}\Pr\{E_j^b\}\Pr\{E_j^l\}\Pr\{E_j^r\} \geq \delta^4 > 0.$$

Step 3: When $E_j$ occurs, we claim that there exists a path $L'(w_\nu, \nu)$ from $w_\nu$ to $\nu$ within $S_j$. If all the paths from $w_\nu$ to $\nu$ go outside $S_j$, they will intersect the closed circuit in $A_j$ and then we can construct a path $L'(w_\nu, \nu)$ within $S_j$ by using part of the closed circuit.

As illustrated in Fig. 7, we place a circle with radius $\frac{1}{2}$ at each user along $L'(w_\nu, \nu)$. It is easy to see that any two circles centered at the two users which are not neighbors on $L'(w_\nu, \nu)$ do not overlap; otherwise we can shorten the path by skipping the users between them. Thus, given the number of hops $|L'(w_\nu, \nu)|$, at least $\left\lceil \frac{|L'(w_\nu, \nu)|}{2} \right\rceil$ nonoverlapping circles centered at alternating nodes on $L'(w_\nu, \nu)$ can be found, and they are all contained within the square with side length $3^j d_\delta + 1$. It follows that $|L'(w_\nu, \nu)| \leq 2 \left\lceil 4(3^j d_\delta + 1)^2/\pi \right\rceil < \infty$, where $2 \left\lceil 4(3^j d_\delta + 1)^2/\pi \right\rceil$ is the maximum number of nonoverlapping circles with radius $\frac{1}{2}$ within the square with side length $3^j d_\delta + 1$.

Step 4: Since $E_j$ are independent, and $\sum_{j=1}^{\infty} \Pr\{E_j\} \geq \sum_{j=1}^{\infty} \delta^4 = \infty$, it follows from the Borel-Cantelli Lemma that $E_j$ occurs for some $j$ a.s. We thus have that $|L'(w_\nu, \nu)| < \infty$ a.s. It implies that $|L(w_\nu, \nu)| \leq |L'(w_\nu, \nu)| < \infty$ a.s., which, together with the a.s. finiteness of the single-hop delay (see Lemma 1), yields the a.s. finiteness of the multihop delay $t^L(w_\nu, \nu)$ along the shortest path $L(w_\nu, \nu)$. Hence, $t(w_\nu, \nu) \leq t^L(w_\nu, \nu) < \infty$ a.s.

## Appendix C: Proof of Lemma 3

The proof is based on the Subadditive Ergodic Theorem which is stated next:

*Fact 1:* [26, Theorem 1.10] Let $\{t_{m,n}\}$ be a collection of random variables indexed by integers satisfying $0 \leq m < n$. Suppose $\{t_{m,n}\}$ has the following properties: (i) $t_{0,n} \leq t_{0,m} + t_{m,n}$; (ii) for each $n$, $\mathbb{E}(|t_{0,n}|) < \infty$ and $\mathbb{E}(t_{0,n}) \geq cn$ for some constant $c > -\infty$; (iii) the distribution of $\{t_{m,m+k} : k \geq 1\}$ does not depend on $m$; (iv) for each $k \geq 1$, $\{t_{nk,(n+1)k} : n \geq 0\}$ is a stationary sequence.

Then: (a) $\eta \triangleq \lim_{n \to \infty} \frac{\mathbb{E}[t_{0,n}]}{n} = \inf_{n \geq 1} \frac{E[t_{0,n}]}{n}$; (b) $T \triangleq \lim_{n \to \infty} \frac{t_{0,n}}{n}$ exists a.s.; (c) $\mathbb{E}[T] = \eta$.

Furthermore, if (v) the stationary sequence in (iv) is ergodic, then (d) $T = \eta$ a.s.

---

[11] Consider two realizations $\omega$ and $\omega'$ of $\mathcal{G}(\lambda_S)$. A partial ordering '$\preceq$' is defined as $\omega \preceq \omega'$ if and only if every node in $\omega$ is also present in $\omega'$. In other words, $\omega$ can be obtained from $\omega'$ by removing some secondary users. An event $E$ is said to be increasing if for every $\omega \preceq \omega'$, $\mathbb{I}_E(\omega) \leq \mathbb{I}_E(\omega')$, where $\mathbb{I}_E$ is the indicator function of the event $E$.

By the definition of the MMD and the stationarity of the CR network model, it is obvious that conditions (i), (iii), and (iv) hold for $\{t_{m,n}\}$. We only need to show that conditions (ii) and (v) also hold for $\{t_{m,n}\}$.

Verification of condition (ii): We first show that $\mathbb{E}(|t_{0,n}|) < \infty$ for each $n$. By using the techniques similar to showing $d(w_\nu, (0,0)) < \infty$ a.s. in the proof of Lemma 2, we can easily see that $d(\tilde{w}_0, (0,0)) < \infty$ a.s. as well as $d(\tilde{w}_n, (n,0)) < \infty$ a.s. It follows that $d(\tilde{w}_0, \tilde{w}_n) \leq d(\tilde{w}_0, (0,0)) + n + d(\tilde{w}_n, (n,0)) < \infty$ a.s.

Let $L(\tilde{w}_0, \tilde{w}_n)$ be the shortest path from $\tilde{w}_0$ to $\tilde{w}_n$. Let $|L|$ denote the number of hops of $L(\tilde{w}_0, \tilde{w}_n)$ and $t_{0,n}^L$ the multihop delay along $L(\tilde{w}_0, \tilde{w}_n)$. Consider the sequence $\{S_j : j \geq 0\}$ of squares constructed in the proof of Lemma 2 (see Fig. 6). For any given $\sqrt[4]{\frac{8}{9}} < \delta < 1$, we choose

$$d = d_\delta \triangleq \max\{\inf\{d' : \Pr\{E_1^u\} \geq \delta \text{ if } d = d'\}, d(\tilde{w}_0, \tilde{w}_n)\}.$$

Similarly, when the event $E_j$ $(j \geq 1)$ occurs, we have $|L| \leq 2 \lceil 4(3^j d_\delta + 1)^2/\pi \rceil$.

If $|L(\tilde{w}_0, \tilde{w}_n)| > 2 \lceil 4(3^j d_\delta + 1)^2/\pi \rceil$, then none of the events $E_1$, $E_2$,...,$E_j$ occur. Thus

$$\Pr\left\{|L| > 2 \lceil 4(3^j d_\delta + 1)^2/\pi \rceil\right\} \leq \prod_{i=1}^{j} \Pr\{E_i^c\} \leq (1 - \delta^4)^j.$$

Let $M = 2 \lceil 4(3d_\delta + 1)^2/\pi \rceil$, then we have

$$
\begin{aligned}
\mathbb{E}[|L|] &= \sum_{k=0}^{\infty} \Pr\{|L| > k\} = \sum_{k=0}^{M} \Pr\{|L| > k\} + \sum_{k=M+1}^{\infty} \Pr\{|L| > k\} \\
&\leq M + \sum_{j=1}^{\infty} 2 \lceil 4(3^{j+1} d_\delta + 1)^2/\pi \rceil \Pr\{|L| > 2 \lceil 4(3^j d_\delta + 1)^2/\pi \rceil\} \\
&\leq M + \sum_{j=1}^{\infty} 2 \lceil 4(3^{j+1} d_\delta + 1)^2/\pi \rceil (1 - \delta^4)^j \\
&\leq M + \frac{72 d_\delta^2}{\pi} \sum_{j=1}^{\infty} 9^j (1 - \delta^4)^j + \frac{48 d_\delta}{\pi} \sum_{j=1}^{\infty} 3^j (1 - \delta^4)^j + 2 \left(\frac{4}{\pi} + 1\right) \sum_{j=1}^{\infty} (1 - \delta^4)^j.
\end{aligned}
$$

If $\delta > \sqrt[4]{\frac{8}{9}}$, $(1 - \delta^4)^j < 9^{-j}$ which implies that $\mathbb{E}[|L|] < \infty$. Let $t_M = \max_{0 \leq d \leq 1} \{\mathbb{E}[t_s(d)]\}$ be the maximum expected single-hop delay for all hop lengths $0 \leq d \leq 1$, then for all $n \geq 1$, $\mathbb{E}[t_{0,n}] \leq \mathbb{E}[t_{0,n}^L] \leq t_M \mathbb{E}[|L|] < \infty$, *i.e.*, $\{t_{m,n}\}$ satisfies the condition (ii).

Verification of condition (v): We show that $\{t_{nk,(n+1)k} : n \geq 0\}$ is mixing[12], which implies its ergodicity. As illustrated in Fig. 8, we construct two squares $S_n$ and $S_{n+j}$ centered at $\left(\frac{(2n+1)k}{2}, 0\right)$ and $\left(\frac{[2(n+j)+1]k}{2}, 0\right)$ with side length $d_n$ and $d_{n+j}$. Let $L_n^*$ be the minimum path from $\tilde{w}_{nk}$ to $\tilde{w}_{(n+1)k}$. We claim that the two minimum paths $L_n^*$ and $L_{n+j}^*$ are a.s. contained in $S_n$ and $S_{n+j}$, respectively, for some $d_n$, $d_{n+j} > 0$. If, for example,

$$\Pr\{E_n\} = \Pr\{L_n^* \text{ is not contained in any finite } S_n\} > 0,$$

---

[12]A measure preserving transformation $T$ is said to be mixing on a probability space $(\Omega, \mathcal{F}, \mu)$ if for all $E, F \in \mathcal{F}$, $\mu(T^n E \cap F) - \mu(E)\mu(F) \to 0$ as $n \to \infty$. A sequence $\{x_k\}$ is said to be mixing if the unit right-shift transformation is mixing on its probability space. The mixing property of a sequence implies its ergodicity [29].

then with a positive probability $|L_n^*| = \infty$, which implies that

$$\mathbb{E}[t_{nk,(n+1)k}] \geq \mathbb{E}[t_{nk,(n+1)k} \mid E_n]\mathrm{Pr}\{E_n\} \geq t_m\mathbb{E}[|L_n^*| \mid E_n]\mathrm{Pr}\{E_n\} = \infty,$$

with $t_m = \min\limits_{0 \leq d \leq 1}\{\mathbb{E}[t_s(d)]\} > 0$ being the minimum expected single-hop delay for all hop lengths[13]. This contradicts $\mathbb{E}[t_{nk,(n+1)k}] < \infty$. Now we have that as $j \to \infty$, not only do the two minimum paths $L_n^*$ and $L^{n+j}$ not share any common secondary users a.s., but also the subsets of the primary transmitter-receiver pairs that affect their multihop delays become disjoint a.s. Thus, $t_{nk,(n+1)k}$ and $t_{(n+j)k,(n+j+1)k}$ are asymptotically independent of each other as $j \to \infty$, $i.e.,$ $\lim\limits_{j \to \infty}\mathrm{Pr}\left\{\left(t_{nk,(n+1)k} < t\right) \cap \left(t_{(n+j)k,(n+j+1)k} < t'\right)\right\} = \mathrm{Pr}\{t_{nk,(n+1)k} < t\}\mathrm{Pr}\{t_{nk,(n+1)k} < t'\}$. The mixing property of $\{t_{nk,(n+1)k} : n \geq 0\}$ follows immediately. Since all the five conditions in Fact 1 are satisfied by $\{t_{m,n}\}$, we conclude that $\exists \beta \geq 0$ such that $\lim\limits_{n \to \infty}\frac{t_{0,n}}{n} = \beta$ a.s.

## Appendix D: Proof of Lemma 4

We will need the following to prove Lemma 4.

*Fact 2:* Given $\mathcal{G}_H(\lambda_S, \lambda_{PT}, t)$ for any $t$ with $(\lambda_S, \lambda_{PT}) \notin \mathcal{C}$, let $B_h = [-h,h]^2$ $(h > 0)$ and take an arbitrary secondary user as the origin. Then $\exists$ $C_1$, $C_2 > 0$ such that $\mathrm{Pr}\{O \leftrightsquigarrow (B_h)^c\} \leq C_1\exp(-C_2 h)$, where $\{O \leftrightsquigarrow (B_h)^c\}$ denotes the event that the origin is connected with some secondary user outside $B_h$, $i.e.,$ the origin and some node in $(B_h)^c$ belong to the same connected component formed by communication links.

This fact can be easily proven by using techniques similar to the ones used in proving Theorem 2.4 in [24]. It provides an upper bound on the CDF of the diameter of the connected component formed by communication links in a secondary network that is not instantaneously connected.

From Fact 1, we know that

$$\beta = \inf_{n \geq 1}\frac{E[t_{0,n}]}{n} \leq \mathbb{E}[t_{0,1}] < \infty.$$

Choose $H > 0$ such that $C_1\exp(-C_2 H) < \frac{1}{2}$, where $C_1$ and $C_2$ are the constants specified in Fact 2. For any path $L$ from $\tilde{w}_0$ to $\tilde{w}_n$, we partition it into several segments in the following way; see Fig. 9. Define a sequence $\{R_i : i \geq 1\}$ of uniformly distributed ribbons on $\mathbb{R}^2$ as

$$R_i = \{(x,y) \in \mathbb{R}^2 : H + (i-1)(H+1) \leq x - x(\tilde{w}_0) < i(H+1)\},$$

where $x(\tilde{w}_0)$ is the x-coordinate of user $\tilde{w}_0$. Since the width of each ribbon is 1 which is equal to $r_p$, there exists at least one user $z_i$ within each $R_i$ that lies between $\tilde{w}_0$ and $\tilde{w}_n$. Assume that these $z_i$ partition the path $L$ into $m$ segments, then the multihop delay $t^L$ along the path $L$ can be written as

$$t^L = \sum_{i=1}^{m} t^L(z_{i-1}, z_i), \tag{D1}$$

where $z_0 = \tilde{w}_0$ and $z_m = \tilde{w}_n$.

---

[13]The inequality $t_m > 0$ is shown in [28, Appendix A].

Based on Fact 2, with a probability greater than $\frac{1}{2}$ at least one hop on the segment of $L$ from $z_{i-1}$ to $z_i$ does not see the opportunity. We thus have that for all $1 \leq i \leq m-1$,

$$\mathbb{E}[t^L(z_{i-1}, z_i)] > \frac{1}{2}t_m, \tag{D2}$$

where $t_m = \min\limits_{0 \leq d \leq 1}\{\mathbb{E}[t(d)]\} > 0$.

Since $d(\tilde{w}_0, \tilde{w}_n) \geq n - d(\tilde{w}_0, (0,0)) - d(\tilde{w}_n, (n,0))$, and both $d(\tilde{w}_0, (0,0))$ and $d(\tilde{w}_n, (n,0))$ are finite a.s., it follows that $\lim\limits_{n \to \infty} \Pr\{d(\tilde{w}_0, \tilde{w}_n) > \frac{n}{2}\} = 1$. When $d(\tilde{w}_0, \tilde{w}_n) > \frac{n}{2}$ holds, any path from $\tilde{w}_0$ to $\tilde{w}_n$ has at least $\left\lfloor \frac{n}{2(H+1)} \right\rfloor$ segments. By recalling (D1), (D2), we conclude that

$$\beta = \lim_{n \to \infty} \frac{\mathbb{E}[t_{0,n}]}{n} > \lim_{n \to \infty} \frac{t_m}{2n}\left\lfloor \frac{n}{2(H+1)} \right\rfloor \Pr\left\{d(\tilde{w}_0, \tilde{w}_n) > \frac{n}{2}\right\} > \lim_{n \to \infty} \frac{t_m}{2}\left(\frac{1}{2(H+1)} - \frac{1}{n}\right) > 0.$$

This implies that $\beta = \lim\limits_{n \to \infty} \frac{t_{0,n}}{n} > 0$.

## Appendix E: Proof of Theorem 3

Similarly to the proof of T2.2, in order to show the a.s. existence of $\lim\limits_{d(\mu,\nu) \to \infty} \frac{t^\tau(\mu,\nu)}{d(\mu,\nu)}$, it suffices to prove that $\lim\limits_{n \to \infty} \frac{t^\tau_{0,n}}{n}$ exists a.s. By the same argument that was used in the proof of Lemma 3, we can easily verify the five conditions in Fact 1. Then the a.s. existence of $\lim\limits_{n \to \infty} \frac{t^\tau_{0,n}}{n}$ follows.

Let $\gamma = \gamma(\tau) = \lim\limits_{d(\mu,\nu) \to \infty} \frac{t^\tau(\mu,\nu)}{d(\mu,\nu)}$. Since the minimum number of hops between $\mu$ and $\nu$ is $\lfloor d(\mu,\nu)/r_p \rfloor$ and the minimum single-hop delay is $\tau$, we have $t^\tau(\mu,\nu) \geq \tau \lfloor d(\mu,\nu)/r_p \rfloor$, which implies that $\gamma \geq \tau/r_p$.

From Fact 1, we have that for any $\tau > 0$,

$$\gamma(\tau) = \lim_{d(\mu,\nu) \to \infty} \frac{t^\tau(\mu,\nu)}{d(\mu,\nu)} = \lim_{d(\mu,\nu) \to \infty} \frac{\mathbb{E}[t^\tau(\mu,\nu)]}{d(\mu,\nu)} \text{ a.s.}$$

Since $\mathbb{E}[t^\tau_{0,n}]$ decreases as $\tau$ decreases and it is strictly positive, it follows that $\lim\limits_{\tau \to 0} \lim\limits_{d(\mu,\nu) \to \infty} \frac{\mathbb{E}[t^\tau(\mu,\nu)]}{d(\mu,\nu)}$ exists. Thus, $\lim\limits_{\tau \to 0} \lim\limits_{d(\mu,\nu) \to \infty} \frac{t^\tau(\mu,\nu)}{d(\mu,\nu)}$ exists a.s.

If $(\lambda_S, \lambda_{PT}) \in \mathcal{C}_I$, then

$$\lim_{\tau \to 0} \lim_{d(\mu,\nu) \to \infty} \frac{t^\tau(\mu,\nu)}{d(\mu,\nu)} \geq \lim_{\tau \to 0} \lim_{d(\mu,\nu) \to \infty} \frac{t(\mu,\nu)}{d(\mu,\nu)} = \beta,$$

where $t(\mu,\nu)$ is the MMD from $\mu$ to $\nu$ when $\tau = 0$, and $\beta$ is defined in (2).

If $(\lambda_S, \lambda_{PT}) \in \mathcal{C}$, then we consider the path $L_C$ from $\mu$ to $\nu$ constructed in the proof of T2.1 which contains some nodes of the infinite connected component $C(t_0)$ in $\mathcal{G}_H(\lambda_S, \lambda_{PT}, t_0)$. Notice that for fixed $d(\mu,\nu)$, only a finite number of hops on $L_C$ belong to $C(t_0)$. Thus if $\tau$ is sufficiently small, it takes at most one primary slot for the message to transmit from the source $\mu$ in $C(t_0)$ to the end node $w_\nu$. Then we have that for some small $\tau_0 = \tau_0(d(\mu,\nu)) > 0$, $t^C_{\tau_0}(\mu,\nu) \leq t_0 + t^{\tau_0}(w_\nu,\nu) + 1$, where $t^C_{\tau_0}(\mu,\nu)$ denotes the multihop delay along the path $L_C$ when the propagation delay is $\tau_0$. It implies that

$$\begin{aligned}
\lim_{\tau \to 0} \gamma(\tau) &= \lim_{\tau \to 0} \lim_{d(\mu,\nu) \to \infty} \frac{\mathbb{E}[t^\tau(\mu,\nu)]}{d(\mu,\nu)} = \lim_{d(\mu,\nu) \to \infty} \lim_{\tau \to 0} \frac{\mathbb{E}[t^\tau(\mu,\nu)]}{d(\mu,\nu)} \\
&\leq \lim_{d(\mu,\nu) \to \infty} \frac{\mathbb{E}[t^C_{\tau_0}(\mu,\nu)]}{d(\mu,\nu)} \leq \lim_{d(\mu,\nu) \to \infty} \frac{\mathbb{E}[t_0] + \mathbb{E}[t^{\tau_0}(w_\nu,\nu)] + 1}{d(\mu,\nu)} = 0,
\end{aligned}$$

since both $\mathbb{E}[t_0]$ and $\mathbb{E}[t^{\tau_0}(w_\nu, \nu)]$ are finite and independent of $d(\mu, \nu)$. In the second equality, we can interchange the order of the two limits because $\mathbb{E}[t^\tau(\mu, \nu)] < \infty$. Consequently, we conclude that a.s.

$$\lim_{\tau \to 0} \lim_{d(\mu, \nu) \to \infty} \frac{t^\tau(\mu, \nu)}{d(\mu, \nu)} = \lim_{\tau \to 0} \gamma(\tau) = 0.$$

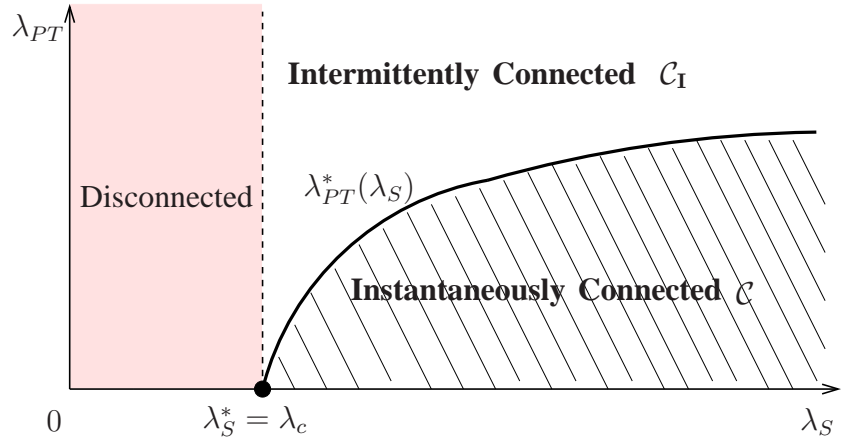

Fig. 1. Connectivity of ad hoc cognitive radio networks (the critical density $\lambda_S^*$ of the secondary users is defined as the infimum density of the secondary users that ensures instantaneous connectivity under a *positive* density of the primary transmitters, and is equal to the critical density $\lambda_c$ of a homogeneous network; the upper boundary $\lambda_{PT}^*(\lambda_S)$ is defined as the supremum density of the primary transmitters that ensures instantaneous connectivity with a *fixed* density of the secondary users).

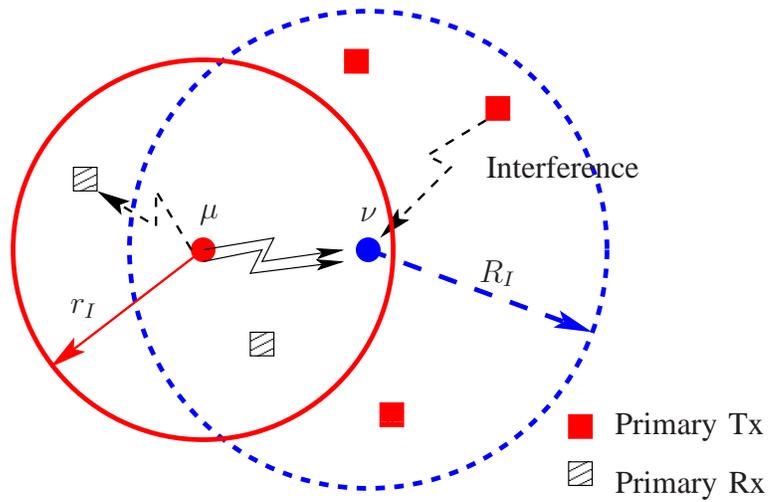

Fig. 2. Definition of spectrum opportunity. $\mu$ and $\nu$ denote secondary transmitter and receiver. $r_I$ and $R_I$ denote the interference radii of the secondary and primary users. A spectrum opportunity from $\mu$ to $\nu$ exists only if there are no primary receivers within the solid circle and no primary transmitters within the dashed circle.

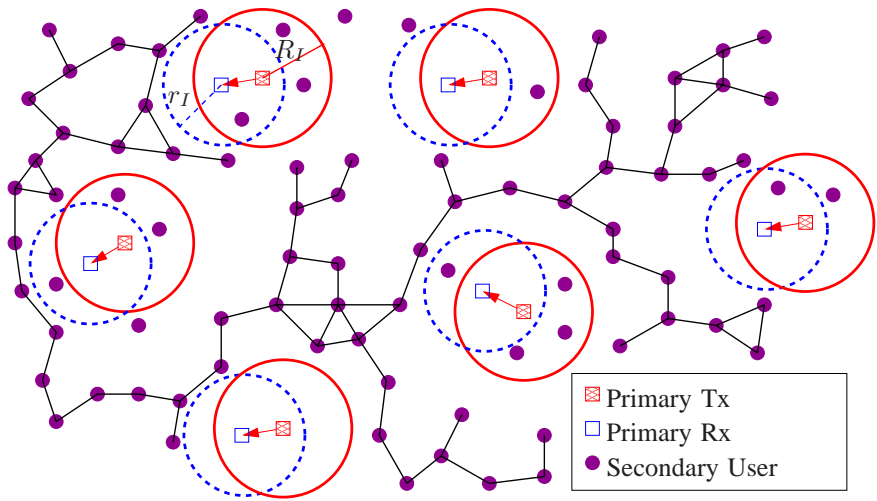

Fig. 3.   A realization of the random graph $\mathcal{G}_H(\lambda_S, \lambda_{PT}, t)$ which consists of all the secondary users and all the communication links in the primary slot $t$ (denoted by solid lines). The solid circles denote the interference regions of the primary transmitters within which secondary users can not successfully receive, and the dashed circles denote the required protection regions for the primary receivers within which secondary users should refrain from transmitting.

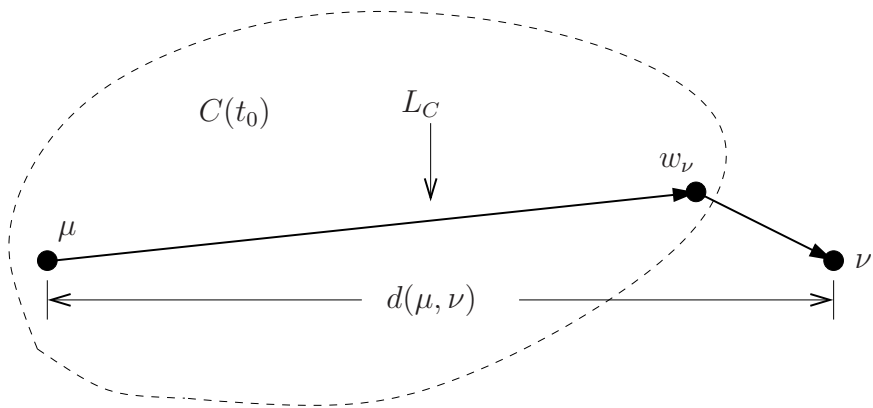

Fig. 4.   An illustration of the constructed path $L_C$ from $\mu$ to $\nu$ when $(\lambda_S, \lambda_{PT}) \in \mathcal{C}$. $C(t_0)$ is the infinite connected component of $\mathcal{G}(\lambda_S, \lambda_{PT}, t_0)$ which first contains $\mu$, and $w_\nu$ is the user in $C(t_0)$ which is closest to $\nu$.

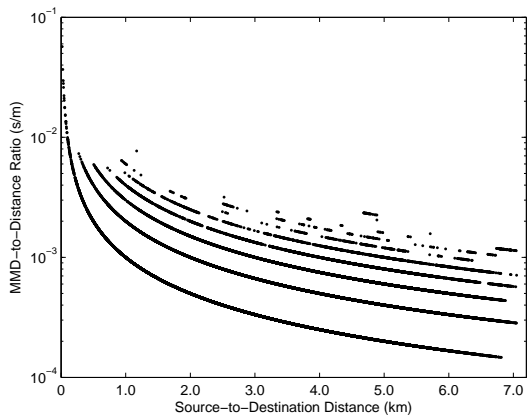

(a) instantaneously connected ($\lambda_{PT} = 10\text{km}^{-2}$, $\tau = 0$)

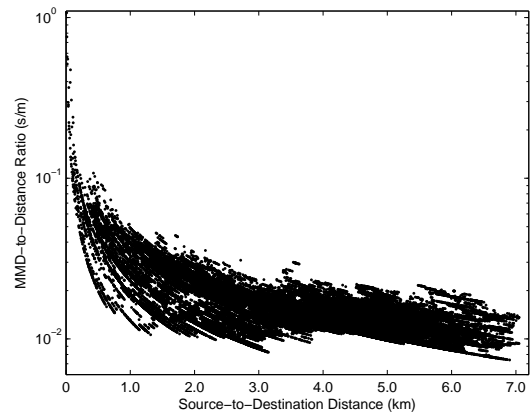

(b) intermittently connected ($\lambda_{PT} = 50\text{km}^{-2}$, $\tau = 0$)

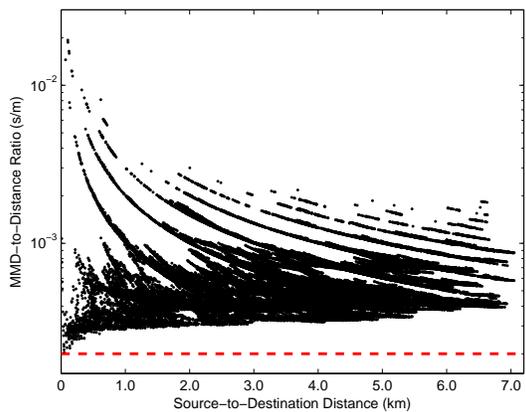

(c) instantaneously connected ($\lambda_{PT} = 10\text{km}^{-2}$, $\tau = 0.01\text{s}$)

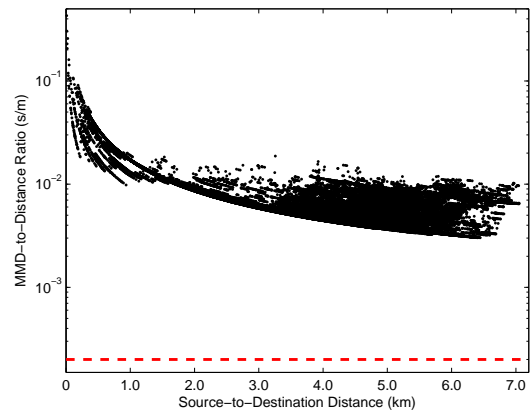

(d) intermittently connected ($\lambda_{PT} = 50\text{km}^{-2}$, $\tau = 0.01\text{s}$)

Fig. 5. MMD-to-distance ratio (in logarithmic scale) vs. the source-to-destination distance. Notice that the MMD-to-distance ratio is obtained in one Monte Carlo run. The secondary users are distributed within a square $[-5\text{km}, 5\text{km}] \times [-5\text{km}, 5\text{km}]$ according to a homogeneous Poisson point process with density $\lambda_S = 700\text{km}^{-2}$. Given the transmission range $r_p = 50m$ of the secondary users, we have that $\lambda_S$ is larger than the critical density $\lambda_c(50) = 576\text{km}^{-2}$. Some other simulation parameters are given by $r_I = 80\text{m}$, $R_p = 50\text{m}$, $R_I = 80\text{m}$, and $T_S = 1\text{s}$.

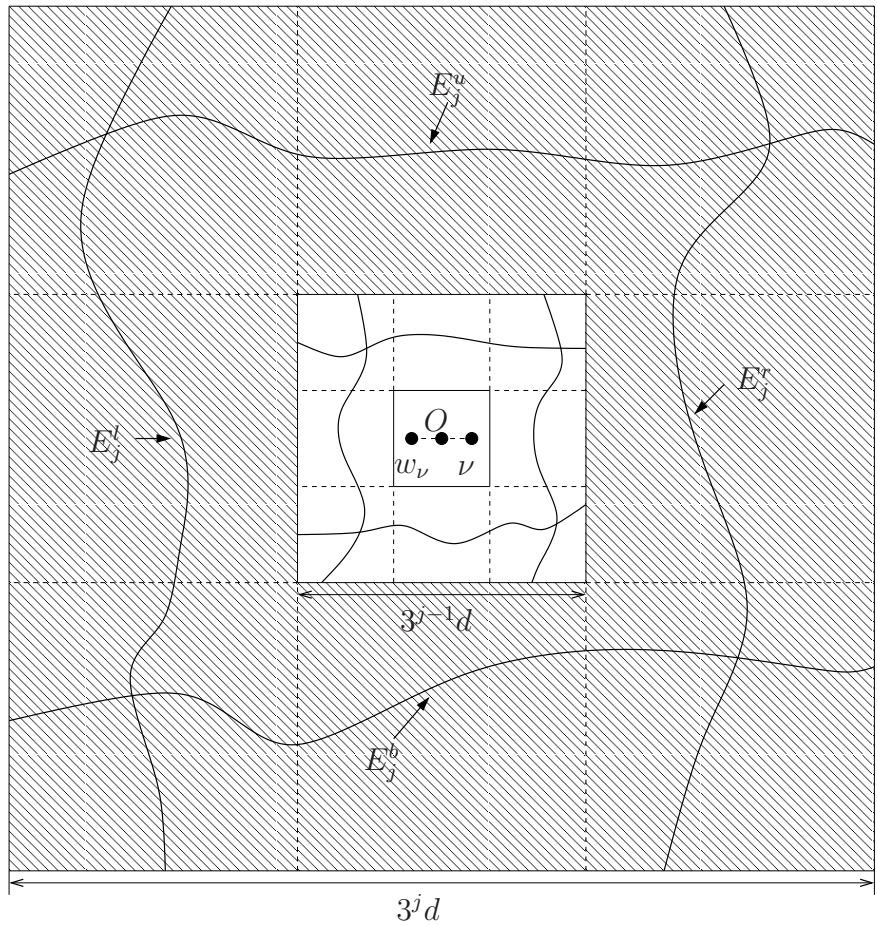

Fig. 6. A sequence $\{S_j : \ j \geq 0\}$ of squares cocentered at the middle point $O$ of $w_\nu$ and $\nu$. The shaded region is the square annulus $A_j$ inside $S_j$ with side length $3^j d$ and outside $S_{j-1}$ with side length $3^{j-1} d$. In this example, the four crossings associated with the four events $E_j^u$, $E_j^b$, $E_j^l$, and $E_j^r$ all exist in the corresponding four rectangles, which form a closed circuit in $A_j$.

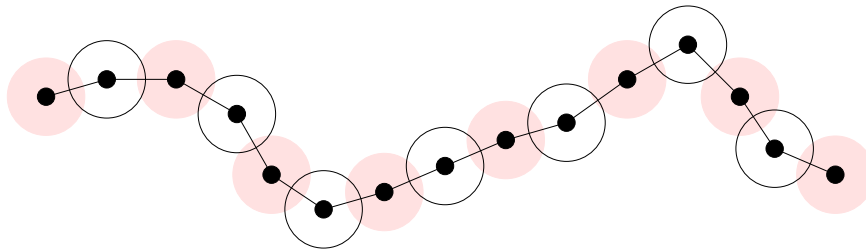

Fig. 7. An illustration of nonoverlapping colored circles with radii $\frac{1}{2}$ centered at alternating nodes on the path $L'(w_\nu, \nu)$.

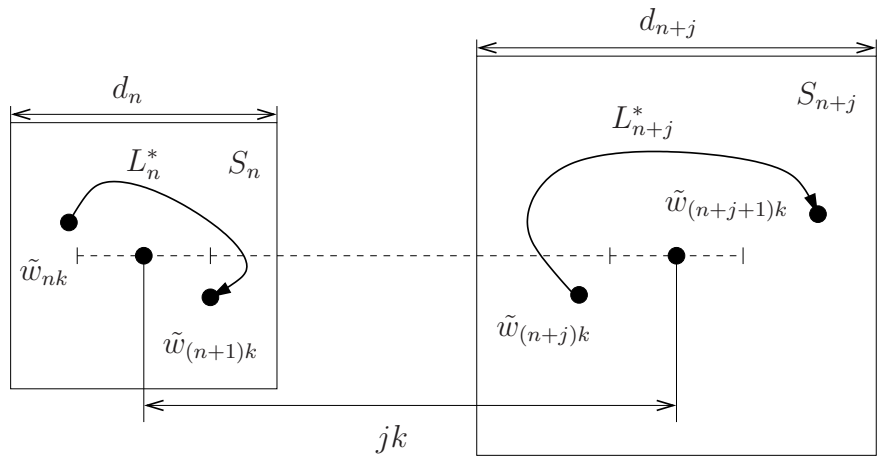

Fig. 8.   The two minimum paths $L_n^*$ (from $\tilde{w}_{nk}$ to $\tilde{w}_{(n+1)k}$) and $L_{n+j}^*$ (from $\tilde{w}_{(n+j)k}$ to $\tilde{w}_{(n+j+1)k}$) are contained in the two squares $S_n$ and $S_{n+j}$ centered at $((2n+1)k/2, 0)$ and $((2n+2\ldots$ PSfrag replacements length $d_n$ and $d_{n+j}$, respectively. As $j \to \infty$, $S_n$ and $S_{n+j}$ become nonoverlapping, and thus the multihop delay along $L_n^*$ is asymptotically independent of the one along $L_{n+j}^*$.

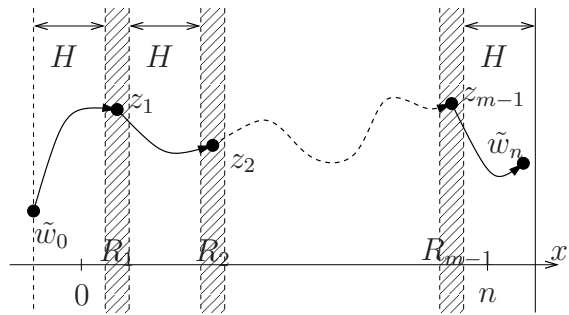

Fig. 9.   A path $L$ from $\tilde{w}_0$ to $\tilde{w}_n$ which is partitioned into $m$ segments by users $z_i$ ($1 \le i \le m-1$) within these shaded ribbons $R_i$ between $\tilde{w}_0$ and $\tilde{w}_n$. Recall that $\tilde{w}_0$ is the user in the infinite topologically connected component which is closest to the coordinate $(0,0)$ and $\tilde{w}_n$ is the user in the infinite topologically connected component which is closest to the coordinate $(n,0)$.